\documentclass[sigconf, nonacm]{acmart}
\setcopyright{none}
\renewcommand\footnotetextcopyrightpermission[1]{}
\usepackage[normalem]{ulem}
\useunder{\uline}{\ul}{}

\begin{document}
\title{An Agentic Software Framework for Data Governance under DPDP}


\author{Apurva Kulkarni}
\orcid{0000-0002-9215-2049}
\affiliation{%
  \institution{International Institute of Information Technology Bangalore}
  \city{Bangalore}
  \state{Karnataka}
  \country{India}
}
\email{apurva.kulkarni@iiitb.ac.in}

\author{Chandrashekar Ramanathan}
\orcid{0000-0002-3330-8365}
\affiliation{%
  \institution{International Institute of Information Technology Bangalore}
  \city{Bangalore}
  \state{Karnataka}
  \country{India}
  }
\email{rc@iiitb.ac.in}

\begin{abstract}
Despite the rise of data-driven software systems in the modern digital landscape, data governance under a legal framework remains a critical challenge. In India, the Digital Personal Data Protection (DPDP) Act mandates rigorous data privacy and compliance requirements, necessitating software frameworks that are both ethical and regulation-aware. 
From a software development perspective, traditional compliance tools often rely on hard-coded rules and static configurations, making them inflexible to dynamic policy updates or evolving legal contexts. Additionally, their monolithic architectures obscure decision-making processes, creating black-box behavior in critical governance workflows. 
Developing responsible AI software demands transparency, traceability, and adaptive enforcement mechanisms that make ethical decisions explainable. To address this challenge, a novel agentic framework is introduced to embed compliance logic directly into software agents that govern and adapt data policies. In this paper, the implementation focuses on the DPDP Act.
The framework integrates \textit{KYU Agent} and \textit{Compliance Agent}  for this purpose. 
\textit{KYU (Know-Your-User) Agent} supports semantic understanding, user trustworthiness modelling  and  \textit{Compliance Agent} uses data sensitivity reasoning within a goal-driven, agentic pipeline. 
The proposed framework, built using an open-sourced agentic framework and has been evaluated across ten diverse domains, including healthcare, education, and e-commerce. Its effectiveness under DPDP, measured via an \textit{Anonymization Score}, demonstrates scalable, compliant data governance through masking, pseudonymization, and generalization strategies tailored to domain-specific needs.
The proposed framework delivers scalable, transparent, and compliant data governance through collaborative agents, dynamic policy enforcement, and domain-aware anonymization.
\end{abstract}


\maketitle
\section{Introduction}
With the rapid proliferation of data-centric systems, organizational data workflows are becoming increasingly complex and heterogeneous. As reliance on data-driven decision-making grows, so does the risk of data breaches, leaks, and fraudulent activities. In response, governments worldwide have introduced stringent data protection regulations, mandating organizations to adopt accountable and transparent data governance practices~\cite{intro}. In the Indian context, the Digital Personal Data Protection (DPDP) Act, 2023 establishes legal requirements for the collection, storage, processing, and transfer of personal digital data.
This regulatory shift underscores the urgent need for robust, explainable, and adaptable data governance frameworks. However, existing solutions often treat governance as a secondary concern, relying primarily on static, rule-based access control mechanisms. Such approaches are insufficient to handle the nuanced legal obligations and evolving contextual demands of modern data ecosystems.
\subsection*{DPDP and Data Governance}
The Digital Personal Data Protection (DPDP) Act\footnote{\url{https://www.meity.gov.in/static/uploads/2024/06/2bf1f0e9f04e6fb4f8fef35e82c42aa5.pdf}} is India’s law for protecting individuals' data and granting individuals the authority to ensure their personal data is collected, stored, and used responsibly. 
It grants individuals rights over their personal data, including the ability to know how it is used, request corrections or deletion, and be notified of any breaches, while establishing clear obligations for organizations and government entities handling such data.

With the rapid growth of data-centric systems, designing solutions that comply with DPDP regulations is critical. From a software engineering perspective, adapting to the DPDP framework requires embedding privacy controls at every stage of the software development lifecycle. Across each phase of the data lifecycle, including collection, storage, sharing, and archival, it is essential to reference DPDP provisions and integrate appropriate technological safeguards. This includes implementing data lineage tracking and comprehensive audit logs to demonstrate compliance, deploying consent management systems to ensure usage aligns with approved purposes, and conducting regular privacy impact assessments. Effective DPDP governance ensures that data is managed with accountability, transparency, and built-in privacy, safeguarding both individual rights and organizational compliance.

\section{Related Work}
In today's data-centric world, privacy is often considered the most important compared to all the other principles of ethical software development. For example, healthcare software that collects sensitive medical data must ensure privacy protections to prevent unauthorized access or misuse. Protecting users' personal information is foundational to building trust and ensuring compliance. However, many frameworks~\cite{sdlc}, such as  Model-View-Controller (MVC) architecture do not have any provision to incorporate privacy into the software design. Similarly, the traditional Software Development Life Cycle (SDLC) addresses only non-functional requirements such as scalability, performance, and availability but without any provision for privacy considerations. In such cases, the software often treats privacy as an add-on or afterthought, typically introducing it in later stages like compliance audits. Any corrections required as a result of non-compliance are either impossible to incorporate or very costly to comply with.
\begin{figure*}
    \centering
    \includegraphics[width=0.85\linewidth, height=12cm]{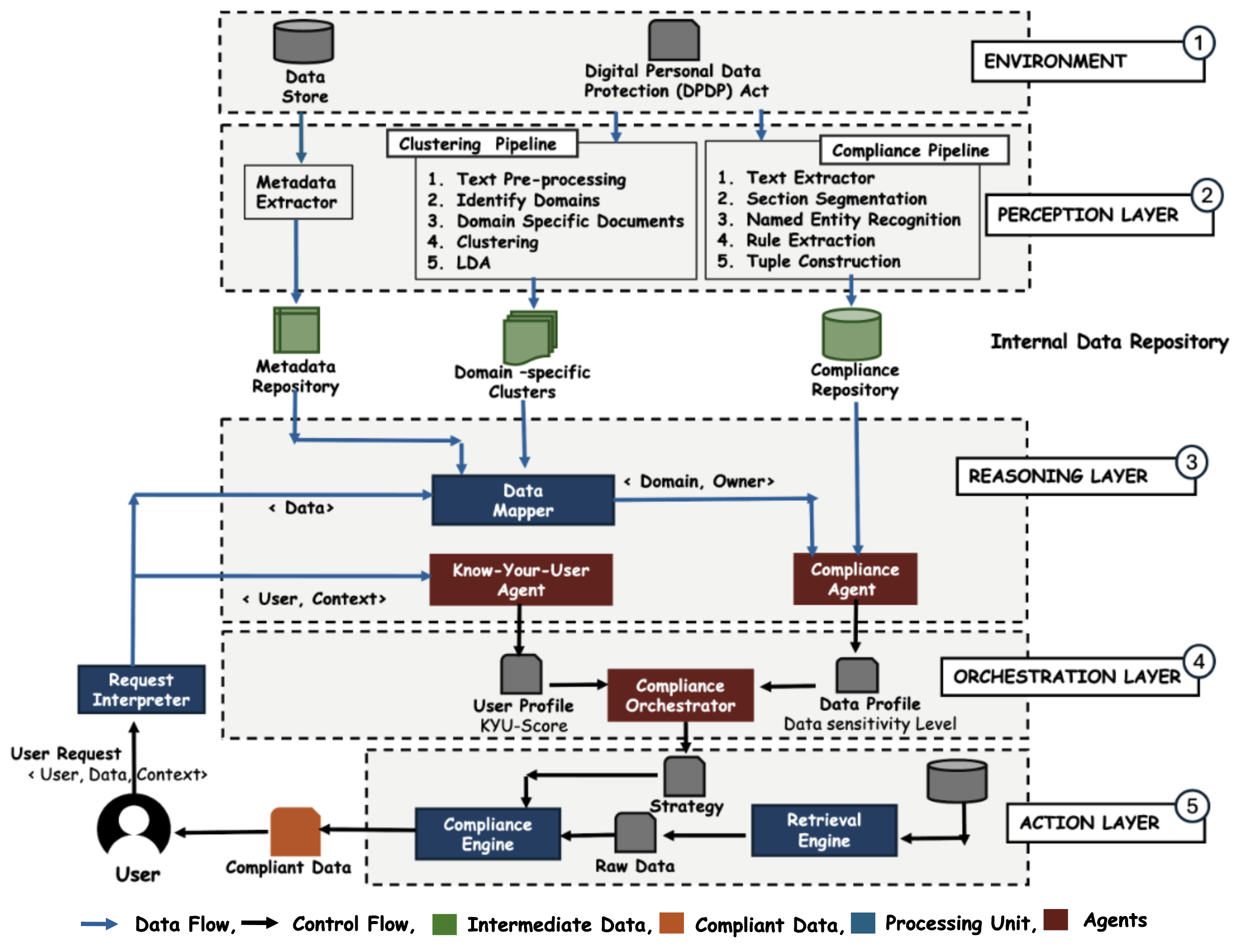}
    \caption{An Agentic Software Framework for Data Governance under DPDP}
    \label{fig:arch}
\end{figure*}
\par Conventional approaches to embedding ethical considerations, such as Privacy by Design~\cite{cavoukian2009privacy}, which integrates privacy measures into the system design from the inception, often conflict with foundational software engineering principles like high cohesion and low coupling. While design paradigms such as Human-Centered Design (HCD)~\cite{hcd1,hcd2}, User-Centered Design (UCD)~\cite{hcd3}, and Value-Sensitive Design (VSD)~\cite{vcd1,vcd2,vcd3} integrate user values, including privacy and accountability, their application remains mostly abstract and suited for greenfield development. Standards like ISO/IEC 27001, Microsoft's Security Development Lifecycle (SDL)~\footnote{\url{https://www.microsoft.com/en-us/securityengineering/sdl/practices}}, and the National Institute of Standards and Technology (NIST)'s  Cybersecurity Framework (CSF)~footnote{\url{https://nvlpubs.nist.gov/nistpubs/CSWP/NIST.CSWP.29.pdf}} provide operational guidance, yet lack integration into software architecture as first-class design constructs.

Furthermore, widely adopted software design models (e.g., Model-View-Controller) and SDLC practices address functional and non-functional requirements but overlook privacy as a systemic, adaptive concern. In practice, this results in privacy being bolted on post-development via tools such as Privacy Impact Assessments (PIAs)\cite{pia}, Data Loss Prevention (DLP) systems\cite{dlp}, and access control frameworks~\cite{acm}, which often respond reactively rather than proactively. Although tools like Open Web Application Security Project (OWASP)~\footnote{\url{https://owasp.org/}}, Zed Attack Proxy (ZAP)~\cite{ozap}, SonarQube~\cite{sonarqube}, and Bandit target security and vulnerability detection, domain-specific privacy solutions like Google Privacy Sandbox~\cite{privacySandbox2024} and ARX~\cite{arx2024} aim at compliance; they remain fragmented and lack a unified, explainable framework. 

\section{Motivation}
Under the Digital Personal Data Protection (DPDP) Act, 2023, which mandates dynamic and purpose-bound data governance, there is a pressing need for \textit{context-aware, agentic frameworks} that operationalize legal semantics directly into software behavior. This work responds to the gap by proposing a modular, domain-agnostic architecture that supports real-time, autonomous governance integrated seamlessly within existing and new data systems.
This work presents a novel agent-oriented software framework that brings automation, compliance, and ethical reasoning into the core of AI-enabled software applications. The primary research contributions are:\\
\textbf{Agent-based Compliance Automation:} Introduces a pluggable, agent-driven framework that operationalizes DPDP compliance as a runtime, context-sensitive decision-making process, enabling dynamic adaptation to user trust levels and data sensitivity.\\
\textbf{Modular Privacy-Preserving Infrastructure:} Develops a software engineering layer for privacy enforcement that is both modular and interoperable, supporting integration with existing and heterogeneous systems via scalable anonymization strategies.\\
\textbf{Quantifiable Cross-Domain Compliance:} Establishes the use of an interpretable, domain-aware \textit{Anonymization Score} as a quantitative software metric for evaluating privacy-preserving effectiveness, enhancing auditability and traceability across sectors.

This framework reframes legal compliance not as a static policy overlay but as a programmable, auditable component of software architecture.
\begin{table*}[h]
\begin{tabular}{|p{3cm}|p{2cm}|p{6cm}|p{3cm}|}
\hline
\multicolumn{1}{|c|}{\textbf{Data Principal}} & \multicolumn{1}{c|}{\textbf{Domain}} & \multicolumn{1}{c|}{\textbf{Rules and Explanation}} & \multicolumn{1}{c|}{\textbf{Receiving Entity}} \\ \hline
Adult Individual & Healthcare & Only with explicit consent. Can be shared with doctors or insurers under Sec 7(f–g) for emergencies or treatment. & Doctors, Insurers  \\ \hline
Child (\textless{}18 years) & Healthcare & Only shared in medical emergencies or by guardian’s consent. No marketing-based sharing. & Guardian, Emergency Services  \\ \hline
Person with Disability (via guardian) & Government Services & Government may process or share data for benefits via the guardian authority. & Guardian, Government \\ \hline
Hindu Undivided Family (HUF) & Finance \& Banking & Shared with tax authorities, legal entities under applicable tax and property laws. & Tax Authorities, Legal Entities  \\ \hline
Company/Firm & Employment \& HR Tech & Employee data can be shared internally for compliance, payroll, and disciplinary action. Not externally without consent/legal order. & Internal Company Departments  \\ \hline
Association or Body of Individuals & Startups and IT Services & Membership data can be shared internally; external sharing must follow Sec 6 consent norms. & Internal Teams  \\ \hline
State & Government Services & May share across departments or contractors under legal authority without user consent. & Government Departments, Contractors  \\ \hline
Artificial Juristic Person (e.g., Trust, NGO) & Startups and IT Services & Can only share personal data of beneficiaries with explicit consent or under governing law. & Legal Authorities, Government Schemes \\ \hline

\end{tabular}
\caption{Selective tuples from Compliance Repository built over DPDP}
\label{tab:tab2}
\end{table*}
\section{Agentic Software Framework for Compliance}
This paper proposes a modular, multi-agent software framework for privacy-aware data access and processing, engineered to align with the DPDP Act. The framework described in Figure~\ref{fig:arch}, adopts a layered, agentic paradigm where distinct functional responsibilities are distributed across perception, reasoning, orchestration, and execution layers. Each layer is designed to operate autonomously while interacting cohesively to ensure that data access decisions are both context-sensitive and legally compliant.
\subsection{Perception Layer}
At the foundational level, the environment layer provides contextual grounding for all agent operations. It incorporates a structured, SQL-based data store that ingests and persists heterogeneous datasets from CSV inputs. Metadata extractor captures the information regarding the data (data type, owner, and domain) and forms a \textit{Metadata Repository.} \\ 
\textbf{Compliance Pipeline:} 
Legal DPDP Act document is written in plain language intended for human understanding, yet its complexity and unstructured nature make it unsuitable for direct machine interpretation. To enable user-specific data handling techniques, sensitivity analysis, and automated compliance reasoning, these laws must be transformed into a structured, machine-interpretable format. The proposed approach develops a \textit{compliance pipeline}, which converts a legal document into a collection of machine-interpretable tuples. \\
The pipeline begins with \texttt{Text Extraction}, where the DPDP Act's unformatted legal text is converted into raw, continuous text while preserving context.
Next, \texttt{Section Segmentation} divides the text into semantically meaningful units such as consent, user rights, data fiduciary obligations, and cross-border data transfer, enabling focused analysis of specific rules.
The segmented text is then processed through \texttt{Named Entity Recognition (NER)} to identify key entities including type of the data (personal, sensitive, anonymized), user roles (Data Principal, Data Fiduciary, Consent Manager), jurisdictions (India, foreign territories), and actions (collection, processing, sharing, erasure). 
Following this, \texttt{Rule Extraction} transforms complex legal language into logically structured, machine-readable rules. These rules are then organized through \texttt{Tuple Construction} into a standard format $\langle$Data Principal, Domain, Rules, Receiving Entity$\rangle$, where the `Data Principal' represents the individual whose data is processed, the `Domain' specifies the relevant sector, the `Rules' encapsulate compliance conditions, and the `Receiving Entity' denotes the party to whom data is disclosed. 

All tuples are compiled into a structured knowledge base, referred to as a  \texttt{Compliance Repository} that is used by the compliance agent to determine data sensitivity.
The current \textit{Compliance Repository} implementation contains 20 derived tuples over the DPDP Act. Table~\ref{tab:tab2} illustrating a selected subset of these entries.\\
\textbf{Clustering Pipeline:} 
The \textit{Clustering Pipeline} adds contextual awareness to the compliance system by determining the domain of each document, ensuring that compliance checks are targeted, relevant, and efficient.\\
It begins with \texttt{Text Pre-processing}, where the corpus (50-60 domain-specific documents) is cleaned using techniques such as tokenization and stop-word removal to prepare it for accurate clustering and downstream NLP tasks. 
In the \texttt{Domain Identification} stage, the content of processed documents is analysed to determine the appropriate domain (e.g., healthcare, finance, education), which is essential because compliance requirements differ across domains. 
The \texttt{Clustering} stage then groups related documents using K-Means clustering, where $k$ equals the number of domains; in our case, 10 distinct domain-specific clusters have been identified~\footnote{The value of $k$ is motivated by the distinct domains mentioned in the DPDP Act}, improving data profiling, retrieval, and compliance matching. To further refine these clusters, Latent Dirichlet Allocation (LDA) is applied to uncover hidden topics within each domain’s documents, enabling organization based on underlying semantic themes rather than surface-level keywords. This is particularly valuable in regulatory contexts where different terms may describe the same concept. 

The output of the \textit{Clustering Pipeline} is the set of \texttt{Domain-Specific Clusters}, which is further leveraged by \textit{Data Mapper} in the \textit{Reasoning Layer}. 

The \textit{Perception Layer} supports both data-centric and policy-centric reasoning and is extensible to accommodate other regulations such as the GDPR (General Data Protection Regulation) or HIPAA (Health Insurance Portability and Accountability Act).
The perception layer transforms raw data and legal text into semantically enriched, structured knowledge through two AI-powered pipelines. 

\subsection{Reasoning Layer}
The reasoning layer supports two AI agents that collaborate to analyze user requests. A request interpreter parses each query into \textit{\(<\) user profile, intent, data type, and access purpose \(>\)}. A data mapper then accesses the \textit{Domain-Specific Cluster} document to identify the relevant cluster and uses a \textit{Metadata Repository} to fetch the owner information.  \\
\textbf{Know-Your-User (KYU) Agent:}
The KYU agent quantifies user trustworthiness using a machine learning model trained to generate a \textit{KYU Score} (trust score). Specifically, a random forest classifier is employed, trained on a synthesized dataset using $k$-fold cross-validation, achieving an accuracy of 98\%. The model evaluates the user's identity and context (as parsed by the request interpreter) to produce a \textit{KYU Score} (trust score) categorized as low, moderate, or high. This score directly informs whether the data can be safely and ethically shared with the user.
\par The model is trained using two key inputs: the requester’s email address and the stated purpose of the request. Email addresses are classified into two trust categories: \textbf{personal} (low trust) and \textbf{organizational} (high trust). Similarly, purposes are categorized into three levels: \textbf{organizational use} (high trust), \textbf{self-use} (moderate trust), and \textbf{external use} (low trust).\\
For example, consider the request for data access with the following attributes: 
\begin{itemize}
    \item \textbf{Email}: \href{mailto:person_1@iiitb.ac.in}{person\_1@iiitb.ac.in}
    \item \textbf{Purpose}: Self Use
    \item \textbf{Requested Attributes}: studentID, Age\_Years, SchoolType
    \item \textbf{Source File}: Education\_Child\_Education.csv
\end{itemize}
In this case, the email domain is organizational (\verb|iiitb.ac.in|), which is assigned a high trust score, while the stated purpose is self-use, which is assigned a moderate trust score. By combining these factors through the KYU Trust Scoring framework, the overall trust score for the request is determined to be \textbf{moderate}, guiding subsequent data access and anonymization strategies.\\
\textbf{Compliance Agent:}
The compliance agent interfaces with the compliance repository to determine the sensitivity of the data being requested. Given the domain and ownership information from the data mapper, the agent identifies matching tuples within the repository. Each tuple’s sensitivity level is computed using a hybrid approach involving large language models (LLMs), specifically LLaMA~\cite{lama} with retrieval-augmented generation (RAG)~\cite{LLMs}. Sensitivity outputs are validated using a human-in-the-loop (HITL) process involving domain experts. The final output is a data sensitivity classification (low, moderate, or high), which contributes to the formation of the data profile used in subsequent policy enforcement.\\
\textbf{Exceptions in Sensitivity Determination:}
If the specific combination of attributes, domain, and ownership provided in a request is not found in any tuple within the compliance repository, the data is by default classified as \textbf{low sensitivity}. This approach aligns with DPDP provisions, where unlisted combinations are not explicitly recognized as protected under the law. Consequently, such data does not trigger heightened safeguards, allowing the compliance agent to apply minimal enforcement measures.
\subsection{Orchestration Layer}
The orchestration layer serves as the policy alignment engine. It synthesizes user trust profiles and data sensitivity classifications to generate context-appropriate privacy strategies. Example of strategies are consent management, anonymization, encryption, so on.  The current implementation employs a rule-based mapping mechanism, wherein the choice of anonymization strategy is governed by the interplay between the \textit{KYU Score} (trust score) and the sensitivity level of the data. For instance, when the \textit{KYU Score} (trust score) is high and data sensitivity is low, the data may be shared in its raw form. Conversely, in scenarios where the \textit{KYU Score} (trust score) is low and data sensitivity is high, stricter anonymization techniques such as masking or encryption are applied to ensure privacy preservation. The output is a concrete privacy action plan referred to as \textit{Strategy}, forwarded to the action layer.
\subsection{Action Layer}
Finally, the action layer operationalizes the \textit{Strategy}. A retrieval engine constructs an SQL query to access the data. The compliance engine then applies the designated privacy-preserving transformation, such as generalization, masking, or encryption, mentioned in the \textit{Strategy}. The policy-compliant output is delivered to the user.

Overall, this architecture represents a software-engineered solution for embedding AI into legally regulated data systems. It advances the design of agentic AI systems that are modular, interpretable, and aligned with data protection laws. By decomposing compliance into discrete, cooperating software agents, the system supports scalable, traceable, and ethically robust data access for engineering responsible AI in real-world applications. 

\section{Evaluation Metric: Anonymisation Score}
To quantitatively assess the degree of privacy preservation under the DPDP Act, we define the \textit{Anonymisation Score} as a normalized measure of transformation applied to sensitive data through masking, generalization, or pseudonymization.
Given a dataset with $N$ records and $M$ attributes, let $O\_{ij}$ and $A\_{ij}$ denote the original and anonymized values, respectively, for the $j$-th attribute in the $i$-th record. The Anonymisation Score is defined as:
\[
\text{Anonymisation Score} = \frac{1}{N \cdot M} \sum_{i=1}^{N} \sum_{j=1}^{M} D(O_{ij}, A_{ij})
\]
where $D(O\_{ij}, A\_{ij})$ is a domain-appropriate context-aware distance metric that quantifies the level of transformation between the original and anonymized values. The score ranges from $0$ (no anonymisation) to $1$ (complete anonymisation), providing a consistent, interpretable basis for evaluating privacy-preserving effectiveness across heterogeneous datasets and domains.\\
The score 0.0 indicates `No Anonymisation', applied when data sensitivity is low and user trust (KYU score) is high; 
0.0 – 1.0 indicates Partial Anonymisation, balances utility and privacy via selective masking or generalization;
1.0 indicates Full Anonymisation, invoked under high sensitivity or low KYU trust scenarios.
Table~\ref{tab:tab11} demonstrates the mean anonymization score captured across the domains. It is observed that domains such as E-commerce, Social Media, Telecom, and Healthcare exhibit higher mean Anonymisation Scores \((\geq 0.62)\), indicating more aggressive privacy enforcement. Sectors like Government, Employment, and Travel, capturing aggregated values, demonstrate lower scores \((\leq 0.40)\), suggesting relatively lenient anonymisation. 
\begin{table}[h]
\centering
\begin{tabular}{l c | l c}
\textbf{Domain} & \textbf{Score} & \textbf{Domain} & \textbf{Score} \\
\hline
E-commerce & 0.63 & Healthcare & 0.62 \\
Social Media & 0.63 & Education & 0.54 \\
Telecom & 0.63 & Finance & 0.49 \\
Startups & 0.44 & Travel & 0.40 \\
Employment & 0.37 & Government & 0.35 \\
\end{tabular}
\caption{Mean Anonymisation Score obtained across domains in experiments}
\label{tab:tab11}
\end{table}
\section{Implementation}
To operationalize privacy-aware decision-making and legal compliance, we employ \textit{CrewAI}, a modular framework for coordinating cooperative intelligent agents in goal-driven workflows. CrewAI enables structured yet flexible orchestration of agents such as the \textit{Compliance Agent}, \textit{KYU Agent}, and \textit{Compliance Orchestrator}. Each agent is defined as a mission-driven component responsible for distinct stages in the data governance pipeline, including metadata resolution, sensitivity scoring, contextual trust estimation, and enforcement via privacy-preserving strategies.

The framework allows agents to operate both independently and collaboratively, supporting rapid prototyping, fault isolation, and rigorous evaluation. This makes \textit{CrewAI} ideal for embedding legal reasoning, policy enforcement, and adaptive anonymisation strategies into real-world data systems. It ensures the architecture remains compliant, explainable, and testable under diverse domains.\\
\textbf{Case Study: Finance Domain}\\
To evaluate the end-to-end system, we ingested 36 datasets spanning multiple sectors. \\
A sample user request from the Finance and Banking domain is as follows:
\\\texttt{\textbf{User Request:}
\\\textbf{Email:} person\_1@gmail.com 
\\\textbf{Purpose:} Organisational Use 
\\\textbf{Requested Attributes:} annual\_income, loan\_status,\\ monthly\_expenditure
\\\textbf{Source File:} Finance\_Banking\_Adult\_FinanceBanking.csv}\\
\par The requested source file is processed by the Data Mapper, which uses \textit{Domain-Specific Clustering} document to classify the domain as `Finance \& Banking' and identify the data owner as `Adult Individual' from the \textit{Metadata Repository}. 
The \textit{KYU Agent} assigns a `Moderate' trust level to the requester after calculating a dynamic trust score based on contextual inputs such as organizational use and a personal email id like person\_1@gmail.com. After receiving this metadata, the \textit{Compliance Agent} evaluates the monthly\_expenditure, loan\_status, and annual\_income data attributes and assigns a sensitivity level as `High' on the basis of relevant tuples from \textit{Compliance Repository}. \textit{Compliance Orchestrator} uses the trust score and sensitivity level to identify  \textit{partial masking anonymization} technique as a strategy for this request. 
\begin{figure}
    \centering
    \includegraphics[width=1\linewidth, height=9cm]{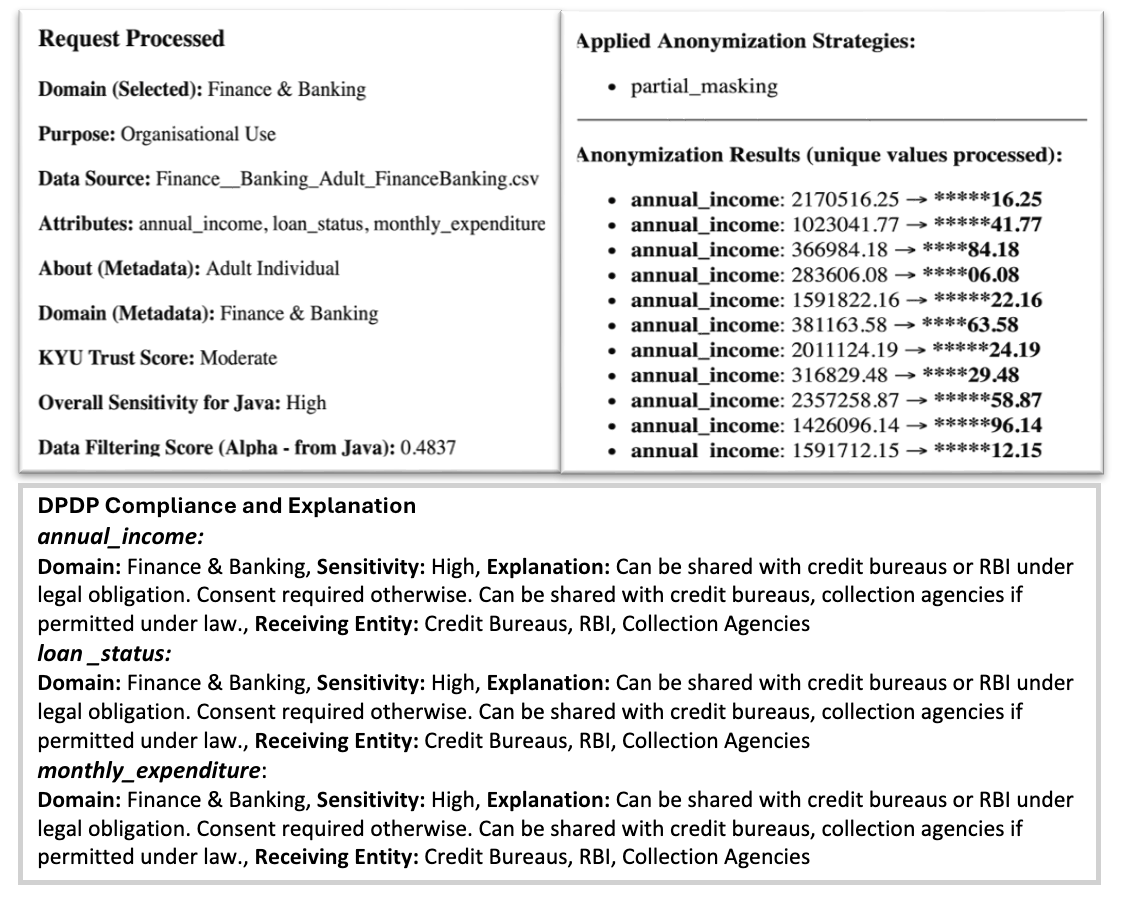}
    \caption{Context-Aware Anonymization and Compliance Justification for Finance \& Banking Data}
    \label{f2}
\end{figure}
Figure~\ref{f2} depicts the output for the user request. With a \textit{moderate} \textit{KYU Score} (trust score) and \textit{high} data sensitivity, the system applied \textit{partial masking}, resulting in an \textit{Anonymisation Score} of \textit{0.4837}.
It also includes legal explanations based on the DPDP Act and keeps clear logs of what changes were made to the data. 
This helps ensure the system is easy to understand, follows the rules, and adapts based on \textbf{who} is asking for the data, \textbf{why} data is being requested (purpose), and \textbf{which} dataset, attributes are requested (owner, domain) — all important parts of building trustworthy and responsible software.

\section{Operationalizing Data Governance}
The proposed approach is aligned with responsible AI principles, emphasizing the flexibility and extensibility of the proposed agentic framework to support key responsible AI features.\\
\textbf{Explainability and Legal Justifiability} are demonstrated through field-level, human-readable justifications rooted in legal norms like the DPDP Act. For each sensitive attribute (e.g., annual\_income), the system outlines its sensitivity, legal basis for sharing, and eligible receiving entities (e.g., RBI, credit bureaus). The dynamic \textit{KYU Score} (trust score) and attribute-level sensitivity scores make anonymization decisions interpretable and auditable in natural language.\\
\textbf{Quantitative Anonymisation Score} bridges AI model behavior and measurable Software Engineering quality. By applying a domain-aware distance metric to quantify transformation, the system produces a normalized score (0–1) for privacy impact. This formal, interpretable metric supports comparison, traceability, and integration into quality assurance pipelines.\\
\textbf{Modular Agent-Oriented Design} using CrewAI allows composable agents like the \textit{Compliance Agent} and \textit{KYU Agent} to manage distinct responsibilities. This modular design supports scalability, maintainability, and testability for orchestrating complex governance workflows.\\
\textbf{Traceability and Auditing} are built-in, with field-level anonymization logs (e.g., 2170516.25 → *****16.25) paired with compliance explanations. Such granular output enables full traceability and auditability, essential for regulated domains.\\
\textbf{Domain Adaptability and Compliance Mapping} allows the system to auto-detect data domains (e.g., Finance) and apply domain-specific compliance rules. This ensures cross-domain scalability and consistent, context-aware enforcement across varied datasets.\\
From a \textbf{software development} standpoint, the framework operationalizes compliance as a first-class software artifact. It formalizes ethical and legal reasoning into programmable entities, enabling machine-interpretable justifications, traceable decision flows, and fine-grained policy enforcement. 
The system doesn’t merely annotate AI decisions post hoc; it governs them at runtime through explainable, testable, and auditable workflows. 
Such automated enforcement pipelines move beyond conventional MLOps to instantiate compliance-aware DevOps for AI systems.

\section{Summary and Future Work}
This work presents an automated, agentic software framework for privacy-preserving data governance, with a focus on dynamic compliance under the DPDP Act. 
By combining contextually enriched processing pipelines, ML-based trust evaluation (KYU Score), and modular agents orchestrated via CrewAI, the system achieves context-aware privacy enforcement. The proposed layered approach of the framework supports maintaining core software engineering principles such as modularity, explainability, and traceability. 
While the current implementation is centered around the DPDP Act, the proposed framework is designed to be adaptable to other regulations and legal frameworks. 
Future work will focus on integrating multiple regulatory requirements into a unified system that can intelligently identify the most relevant context and resolve any conflicts among them. Additionally, the framework can be extended to support cross-border governance and compliance scenarios. From a technical standpoint, future work involves incorporating Retrieval-Augmented Generation (RAG) to enhance contextual understanding and interpretation of legal texts.
These enhancements aim to push the boundary of compliance-by-design and ethics-as-code, positioning the framework as a foundational step toward fully automated, ethically aligned software systems.
\section*{Acknowledgments}
\noindent The research activities described in the paper were supported by (a) Center for Internet of Ethical Things established by the Karnataka Innovation \& Technology Society, Dept. of IT, BT and S\&T, Government of Karnataka, India and (b) Center for Technology Research and Innovation (Digital Governance) established by the Center for E-Governance, Government of Karnataka, India.

\bibliographystyle{ACM-Reference-Format}
\bibliography{references}

\end{document}